\documentclass[letterpaper]{article} 
\usepackage[margin=2cm]{geometry}
\usepackage{times} 
\usepackage[hyphens]{url}
\usepackage{graphicx}
\usepackage{apacite}
\newcommand{\citet}[1]{\citeA{#1}}
\usepackage{caption}
\usepackage{todonotes}
\usepackage[font={small}]{subfig}
\usepackage{booktabs} 
\usepackage{multirow} 

\title{Tiplines to Combat Misinformation on Encrypted Platforms:\\A Case Study of the 2019 Indian Election on WhatsApp}

\author {
    Ashkan Kazemi,\textsuperscript{1,4}
    Kiran Garimella,\textsuperscript{2}
    Gautam Kishore Shahi,\textsuperscript{3} \\
    Devin Gaffney,\textsuperscript{4}
    Scott A.\ Hale\textsuperscript{4,5}%
\\{\small
    \textsuperscript{1} University of Michigan,
    \textsuperscript{2} MIT,
    \textsuperscript{3} University of Duisburg-Essen, 
    \textsuperscript{4} Meedan,
    \textsuperscript{5} University of Oxford} 
\\{\small ashkank@umich.edu, garimell@mit.edu, gautam.shahi@uni-due.de, devin@meedan.com, scott@meedan.com}%
}

\date{}

\usepackage{fancyhdr}
\begin{document}

\maketitle

There is currently no easy way to fact-check content on WhatsApp and other end-to-end encrypted platforms at scale.
In this paper, we analyze the usefulness of a crowd-sourced  ``tipline'' through which users can submit content (`tips') that they want fact-checked.
We compare the tips sent to a WhatsApp tipline run during the 2019 Indian national elections with the messages circulating in large, public groups on WhatsApp and other social media platforms during the same period.
We find that tiplines are a very useful lens into WhatsApp conversations: a significant fraction of messages and images sent to the tipline match with the content being shared on public WhatsApp groups and other social media.
Our analysis also shows that tiplines cover the most popular content well, and a majority of such content is often shared to the tipline before appearing in large, public WhatsApp groups. Overall, our findings suggest tiplines can be an effective source for discovering content to fact-check. 

\section{Research Questions}
\begin{itemize}
    \item How effective are tiplines for fact-checking encrypted social media?
    \item What content is submitted to tiplines for fact-checking?
\end{itemize}

\section{Essay Summary}
\begin{itemize}
    \item A \textit{tipline} is a dedicated service to which messages (`tips') can be submitted by users. On WhatsApp, a tipline would be a phone number to which WhatsApp users can forward information they see, in order to have it fact-checked.
    
    \item Using state-of-the-art text and image matching techniques, we compare content sent to the tipline to the content collected from a large-scale crawl of public WhatsApp groups, ShareChat (a popular image sharing platform similar to Instagram in India), and fact-checks published during the same time in order to understand the overlap between these sources.
    
    \item The tipline covers a significant portion of popular content with 67\% of popular images and 23\% of popular text messages shared on public WhatsApp groups also appearing on the tipline.
    
    \item We find that a majority of the `viral' content spreading on WhatsApp public groups and on ShareChat is shared on the WhatsApp tipline first.

    \item Compared to content by popular fact-checking organizations, the coverage of WhatsApp public group messages provided by the tipline is much higher. We suspect this is because fact-checking organizations typically fact-check content primarily based on signals from open social media platforms like Facebook and Twitter, whereas the tipline is a crowdsourced collection of content native to WhatsApp.
\end{itemize}

\section{Implications}

\begin{table}[t]
\small
\caption{Examples of text messages forwarded to the WhatsApp tipline to be fact-checked (translated to English). The full content submitted includes a variety of topics and content types (e.g., text, images, links, videos and audio).}
\label{table:example}
\begin{center}
\begin{tabular}{ p{0.95\columnwidth}} 
\toprule
 UNESCO Declare India’s ``Jana Gana Mana'' the World’s Best National Anthem \\
\midrule
 When you reach poling booth and find that your name is not in voter list, just show your Aadhar card or voter ID and ask for ``challenge vote'' under section 49A and cast your vote.
 If you find that someone has already cast your vote, then ask for ``tender vote'' and cast your vote.
If any polling booth records more than 14\% tender votes, repolling will be conducted in such poling booth.
 Please share this very important message with maximum groups and friends as everyone should aware of their right to vote. \\
\midrule
 Happened today on 47 street (Diamond Market) New York \$100,000 given away in ref to Modi victory .. see how this millionaire Indian is doing .. \\
\midrule
Coal India is on the verge of ruin! 85,000 crore loss due to Modi!
\url{https://boltahindustan.in/bh-news/coal-india -faced-lost-of-85000-crores/}  \\
\bottomrule
\end{tabular}

\end{center}
\end{table} 

Platforms such as WhatsApp that offer end-to-end encrypted messaging face challenges in applying existing content moderation methodologies since only users involved in the communications have access to the content.
Even though WhatsApp is extremely popular, used by over 2 billion users all over the world, there is currently no large-scale way to understand and debunk misinformation spreading on the platform. Given the real life consequences of misinformation~\cite{arun2019whatsapp} and the increasing number of end-to-end encrypted platforms, developing tools to understand and combat misinformation on these platforms is a pressing concern. 


One potential solution is to make use of ``misinformation tiplines'' \cite{meedan2020oneyear} for checking content veracity and ensuring platform health. 
A \textit{tipline} is a dedicated service to which `tips' can be submitted by users. On WhatsApp, a tipline would be a phone number to which WhatsApp users can forward information they see, in order to have it fact-checked. 
Tiplines represent an example of Distributed Human Computation \cite{yang2011distributed} to discover misinformation while preserving user privacy.

In this study, we use data from a tipline during the 2019 Indian elections, called the CheckPoint project.\footnote{\url{https://www.checkpoint.pro.to/}} 
Checkpoint was a research project led by PROTO and Pop-Up Newsroom, technically assisted by WhatsApp.\footnote{Pop-Up Newsroom is a joint project of Meedan and Fathm, that designs and leads global election and event monitoring journalism efforts.} The goal of this project was to study the misinformation phenomenon at scale---natively in WhatsApp---during the Indian elections.
The tipline was advertised in the national and international press during the elections.\footnote{\url{https://techcrunch.com/2019/04/02/whatsapp-adds-a-tip-line-for-checking-fakes-in-india-ahead-of-elections/}}
Table \ref{table:example} presents some examples of text messages submitted to the tipline. Our goal is to understand what content is submitted, analyze how effective tiplines can be for fact-checking efforts, and shed light on the otherwise black-box nature of content spreading on WhatsApp. 

Our results show the effectiveness of tiplines in content discovery for fact-checking on encrypted platforms. We show that: 
\begin{enumerate}
    \item A majority of the `viral' content spreading on WhatsApp public groups and on ShareChat is shared on the WhatsApp tipline first, which is important as early identification is an essential element of an effective fact-checking pipeline given how fast rumors can spread \cite{vosoughi2018spread}.
    
    \item The tipline covers a significant portion of popular content with 67\% of popular images and 23\% of popular text messages shared on public WhatsApp groups also appearing on the tipline. 

    \item When compared to content by popular fact-checking organizations, the coverage of WhatsApp public group messages provided by the tipline is much higher. We suspect this is because fact-checking organizations typically fact-check content primarily based on signals from open social media platforms like Facebook and Twitter, whereas the tipline is a crowdsourced collection of content native to WhatsApp.
\end{enumerate}

These insights demonstrate tiplines can be an effective privacy-preserving, opt-in solution for fact-checking and combating misinformation on WhatsApp and other end-to-end encrypted platforms.
%
%
There are three main stakeholders who could benefit from this research: academics, fact-checking organizations, and social media companies.

\noindent\textbf{Academics and fact-checking organizations}.
Researchers or journalists trying to use data from encrypted social media apps like WhatsApp could make use of data from such tiplines to study WhatsApp.

The current model to identify and fact check viral content on WhatsApp is to monitor conversations in a convenience sample of public WhatsApp groups~\cite{garimella2020images,melo2019whatsappmonitor}. However, this is non-trivial and resource intensive to manage.
Another solution that fact-checking organizations follow is to just monitor non-encrypted social media platforms such as Facebook or Twitter and assume that content viral on one of these platforms likely overlaps with viral content on other platforms. 

Our work shows that there are far more matches between tipline content and public group messages on WhatsApp than between public group messages and either published fact-checks or open social media content. This notable difference in the content coverage found in tiplines versus fact-checked content stresses the opportunity tiplines provide in identifying and countering misinformation in encrypted platforms.
Even with less than 10\% of the amount of content shared on public WhatsApp groups, our analysis showes tiplines can effectively help monitor and fact-check the most viral content.
They provide high-quality content that is most likely fact-check worthy.


Our analysis also found that most users submitting to the tipline were quite involved and motivated to have their content fact checked: users would often follow up on content they submitted if it had not yet been fact-checked. 
In order to manage a large tipline at scale, fact-checking organizations and academics will need to partner together to create and use technology~\cite{kazemi2021claim,konstantinovskiy2018towards} and create meaningful tipline experiences for users. This will involve setup costs and take time to foster dedicated contributors who are willing to forward potentially misleading content to a tipline.


%

It's worth noting that the tipline, public group, and fact-check content we study were drawn from a specific time period around a large political event (the 2019 Indian Elections). It is unclear how the dynamics may differ for a less eventful time period. Several always-on WhatsApp misinformation tiplines launched in December 2019 and the number has grown since. We encourage researchers to support civil society organizations running these tiplines as they represent a valuable source to better understand the dynamics of misinformation on such end-to-end encrypted platforms.

\noindent\textbf{Social media companies}.
Tiplines are also one of the only sources for encrypted social media platforms to understand harmful content spreading on their platform and take action.
Tiplines can be used to collect `signatures' of popular misleading or hateful content, which in turn can be used to develop on-device solutions which work in encrypted settings. 
For instance, \citet{reis2020ondevice} examine images and propose an on-device approach to alerting users to content that has been fact-checked on WhatsApp. Their solution focuses on PDQ hashes for images and requires a list of hashes for known pieces of misinformation. Our analysis in this paper suggests that tiplines could be a successful way to populate such a hashlist. The most popular images are likely to be submitted to a tipline, and, even better, they are very likely to be submitted to the tipline before they are widely shared within public groups. Thus if a hashlist was populated based on images sent to tiplines, it would identify a large number of these shares.

Using advances in the state-of-the-art techniques to find similar image and text messages, an on-device fact-checking solution could identify up to 40\% of the shares of misinformation in public WhatsApp groups while preserving end-to-end encryption as explained further when addressing research question 1.


\section{Findings}

\subsection{Research Question \#1}
\label{sec:tipline_effectiveness}

To study the effectiveness of tiplines as a source of content for fact-checking, we compare the tipline content to other sources of data, including public groups, fact-checks, and social media.
Our analysis is primarily along two lines: Firstly, how long does it take for an item to be shared by someone to the tipline? 
The intuition behind this is to check how long it takes for tiplines to surface content spreading in public groups and on social media.
Second, what is the coverage of the types of content shared on the tipline in terms of popularity? We want to understand how the probability of content being submitted to the tipline changes with its popularity.

\begin{figure} 
    \centering
    \includegraphics[width=0.5\textwidth]{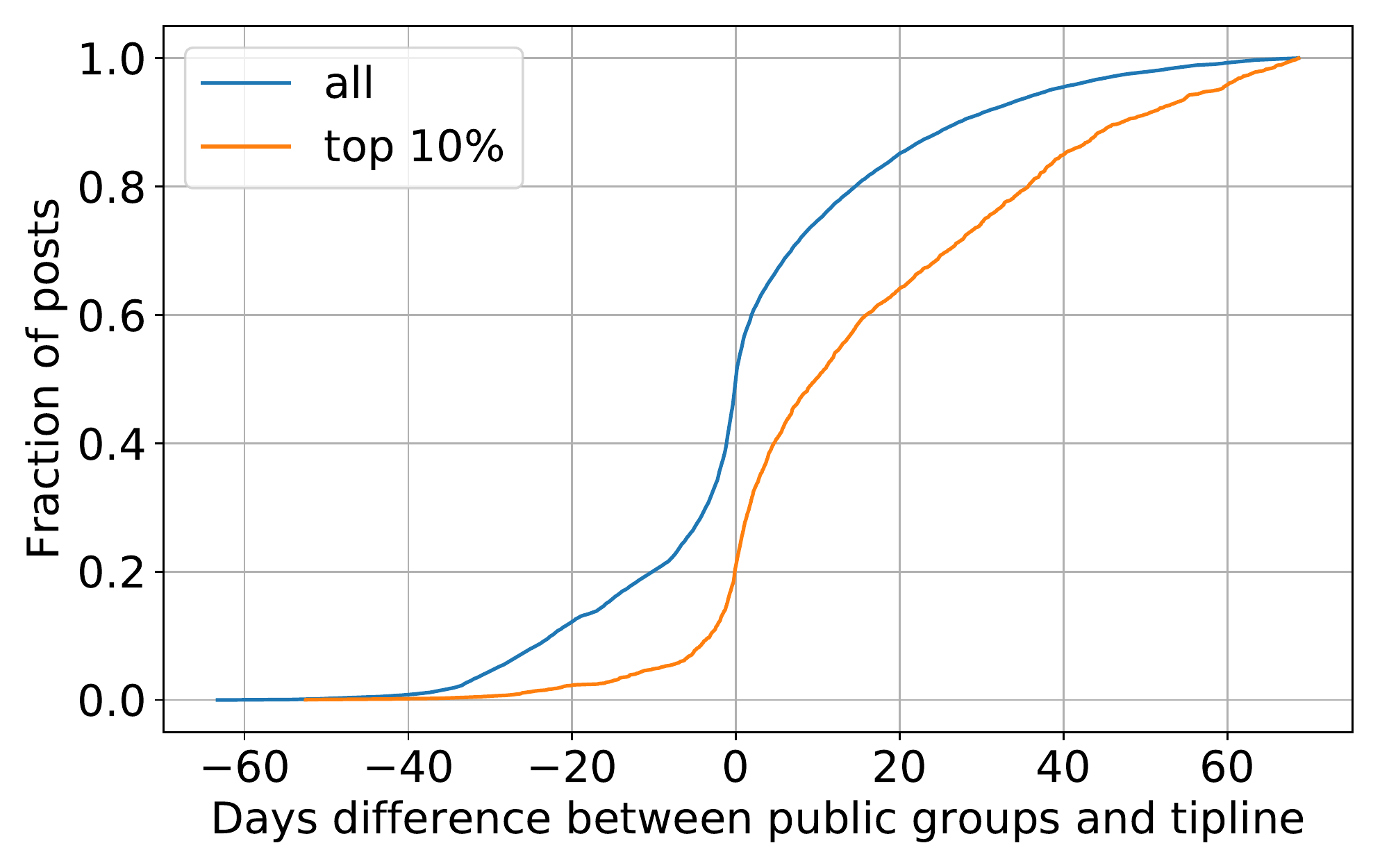}
    \caption{The CDF of the time difference between the sharing of images on public groups and the tipline. Approximately 50\% of the images (blue line) were shared on public groups first. However, if we consider just the top 10\% most shared images in the public groups, they are mostly shared first on the tipline.}
    \label{fig:img-checkpoint-timediff}
\end{figure}

\begin{figure}
    \centering
    \includegraphics[width=0.5\textwidth]{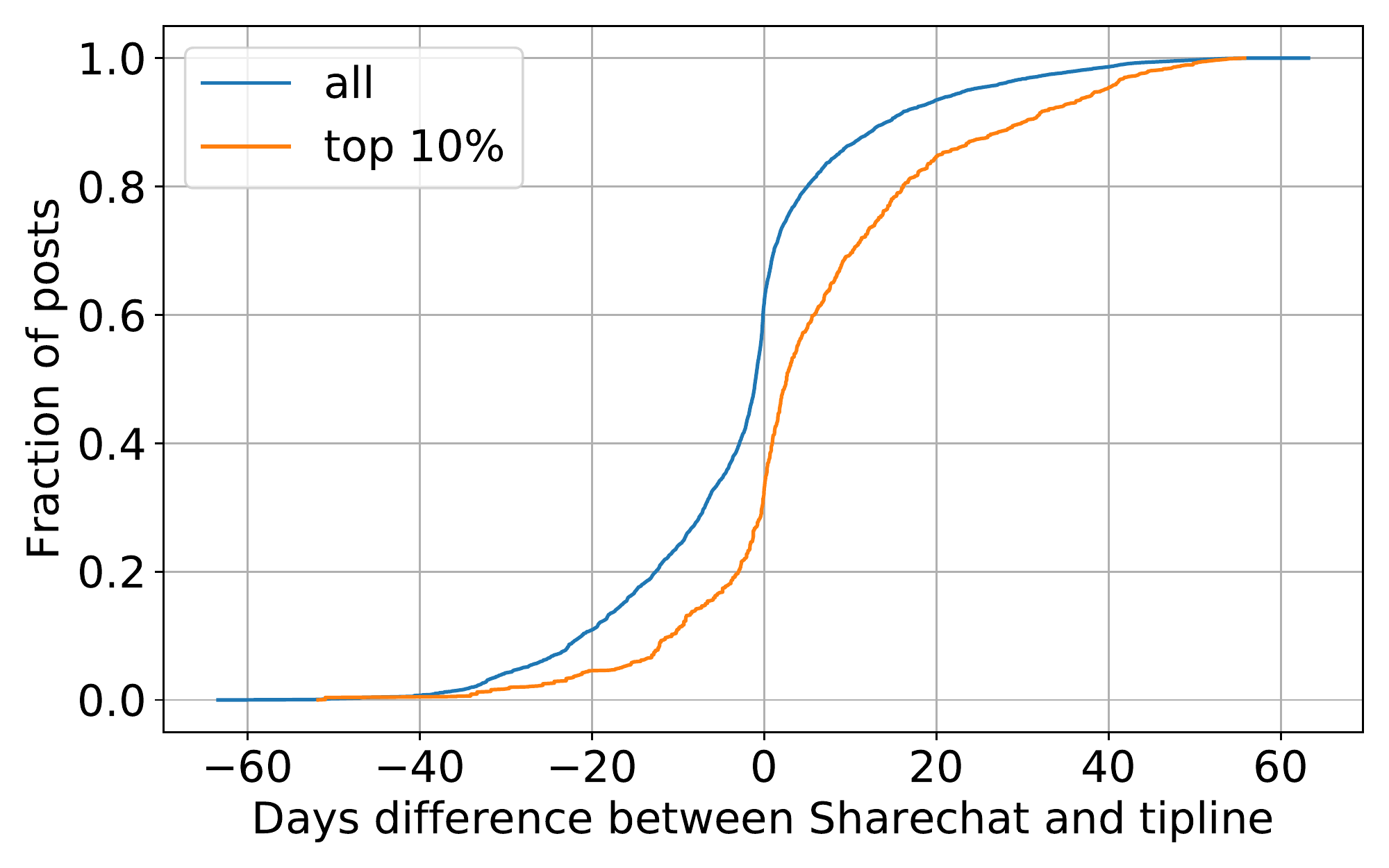}
    \caption{Time difference for images shared on ShareChat and the tipline. The most popular content is more likely to be shared on tiplines first compared to all content.} 
    \label{fig:img-checkpoint-sharechat-timediff}
\end{figure}

Figure~\ref{fig:img-checkpoint-timediff} shows the time difference between an image being shared on a public group and the tipline. Negative values on the x-axis indicate that the content is shared in a public group first.
We see that roughly 50\% of all the content is shared in public groups first, with around 10\% of content going back to over a month.
However, if we focus on the subset of top-10\% most shared images within the public groups, the distribution looks very different. We clearly see that a majority of the content (around 80\%) is first shared on the tipline, indicating that the tipline does a good job covering the most popular content quickly.
Similar trends exist for images on ShareChat (Figure~\ref{fig:img-checkpoint-sharechat-timediff}). In fact, images sent to the tipline have significantly more shares (41 vs. 29) and likes (51 vs. 40) on ShareChat compared to images not sent to the tipline ($p<0.01$ for a t-test of means).

\begin{figure}
    \centering
    \includegraphics[width=0.5\textwidth]{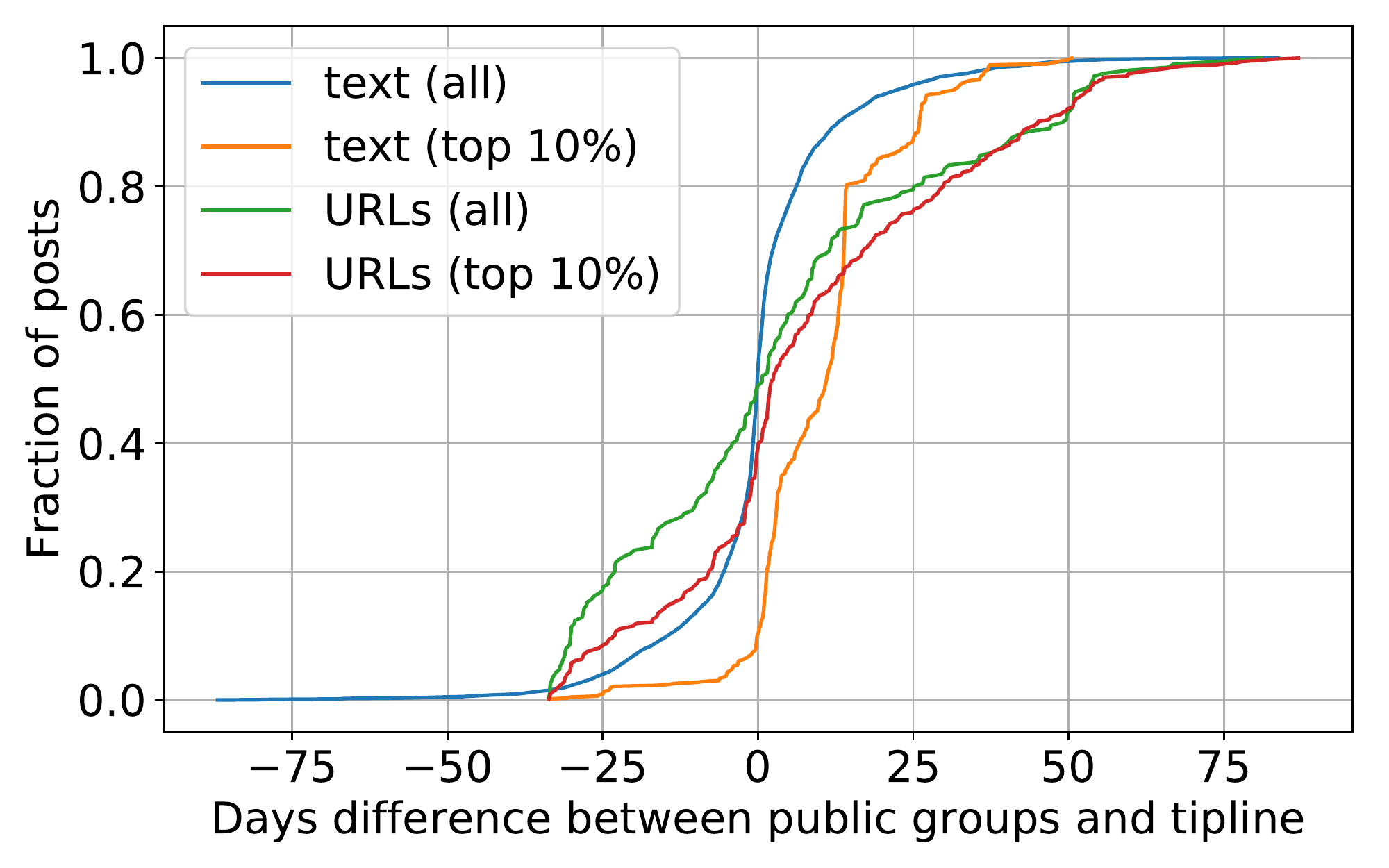}
    \caption{Time difference between the sharing of text messages and URLs in the WhatsApp tipline and public groups.} 
    \label{fig:text-checkpoint-whatsapp-timediff}
\end{figure}

Comparing the text messages within the public groups to the tipline messages leads to similar results (Figure~\ref{fig:text-checkpoint-whatsapp-timediff}). 
To make this comparison, we first clustered all text messages in the public groups and, separately, in the tipline. 
This comparison only uses the text messages from the tipline within clusters having at least five unique messages that were annotated as having claims that could be fact-checked to avoid the risk of match spam or less meaningful content. We again find that the most popular content is often shared first to the tiplines before spreading widely within public groups.
Similar trends also exist for URLs (Figure~\ref{fig:text-checkpoint-whatsapp-timediff}, green and red lines).

\begin{figure} 
    \centering
    \includegraphics[width=0.5\textwidth]{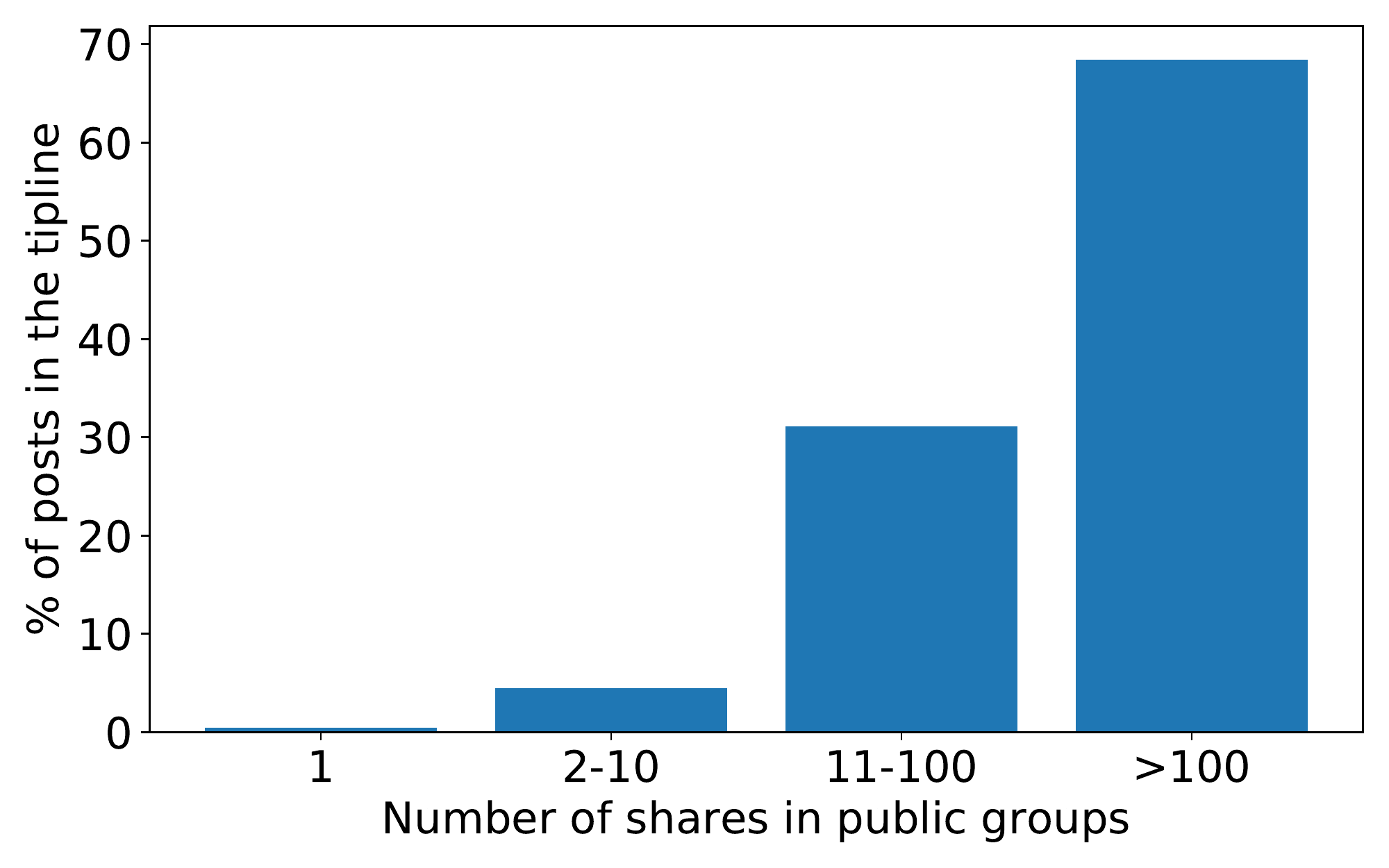}
    \caption{Coverage of Images: The x-axis shows the number of shares on the public groups and y-axis shows the percentage of images with x shares that match with an image submitted to the tipline. Images that are highly shared on the public groups are much more likely to be also shared to the tipline.}
    \label{fig:img-checkpoint-coverage}
\end{figure}

\begin{figure}
    \centering
    \includegraphics[width=0.5\textwidth]{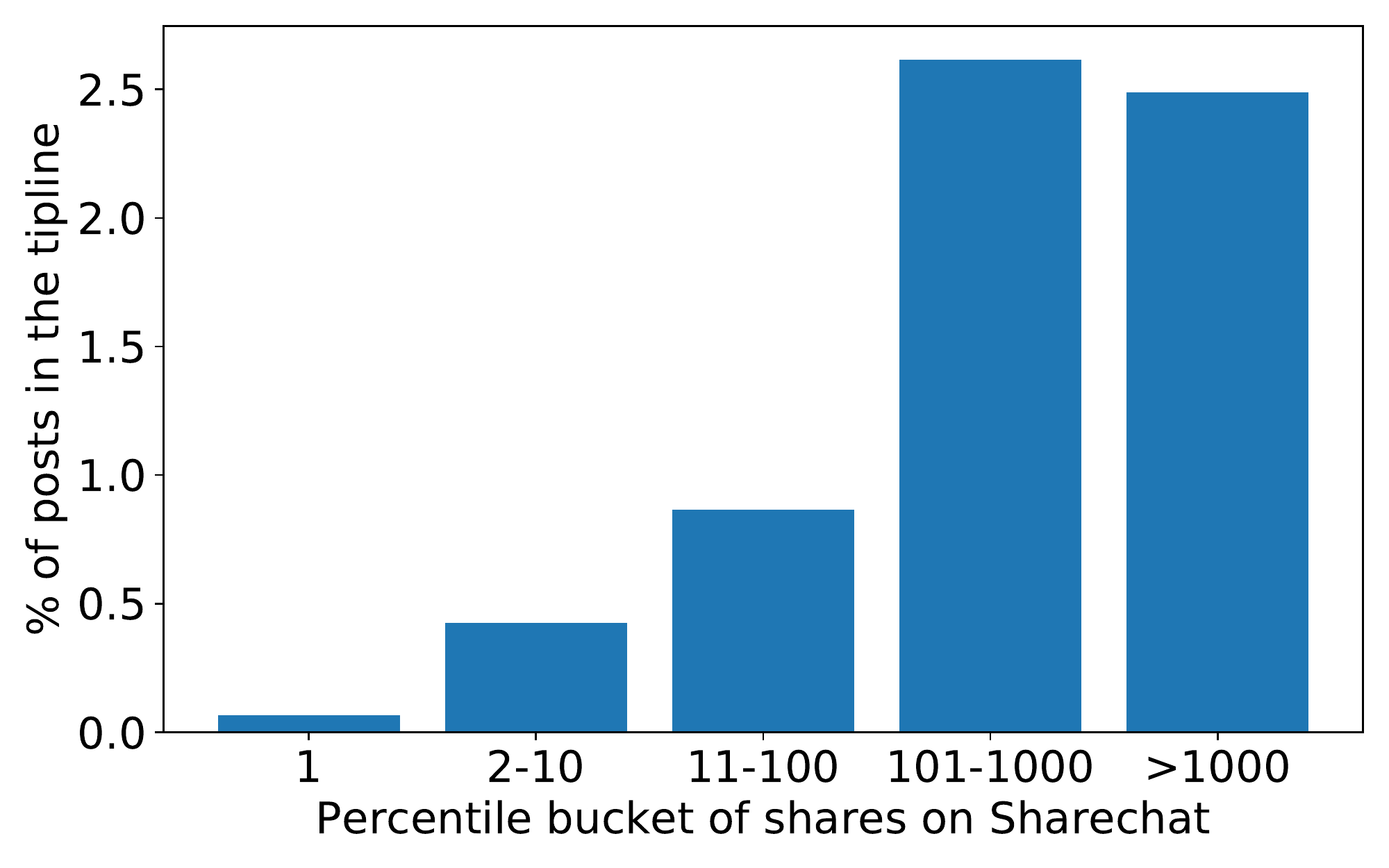}
    \caption{Coverage: Similar to Figure~\protect\ref{fig:img-checkpoint-coverage}, images shared more often on ShareChat are more likely to appear in the tipline.} 
    \label{fig:img-checkpoint-sharechat-coverage}
\end{figure}

\begin{figure}
    \centering
    \includegraphics[width=0.5\textwidth]{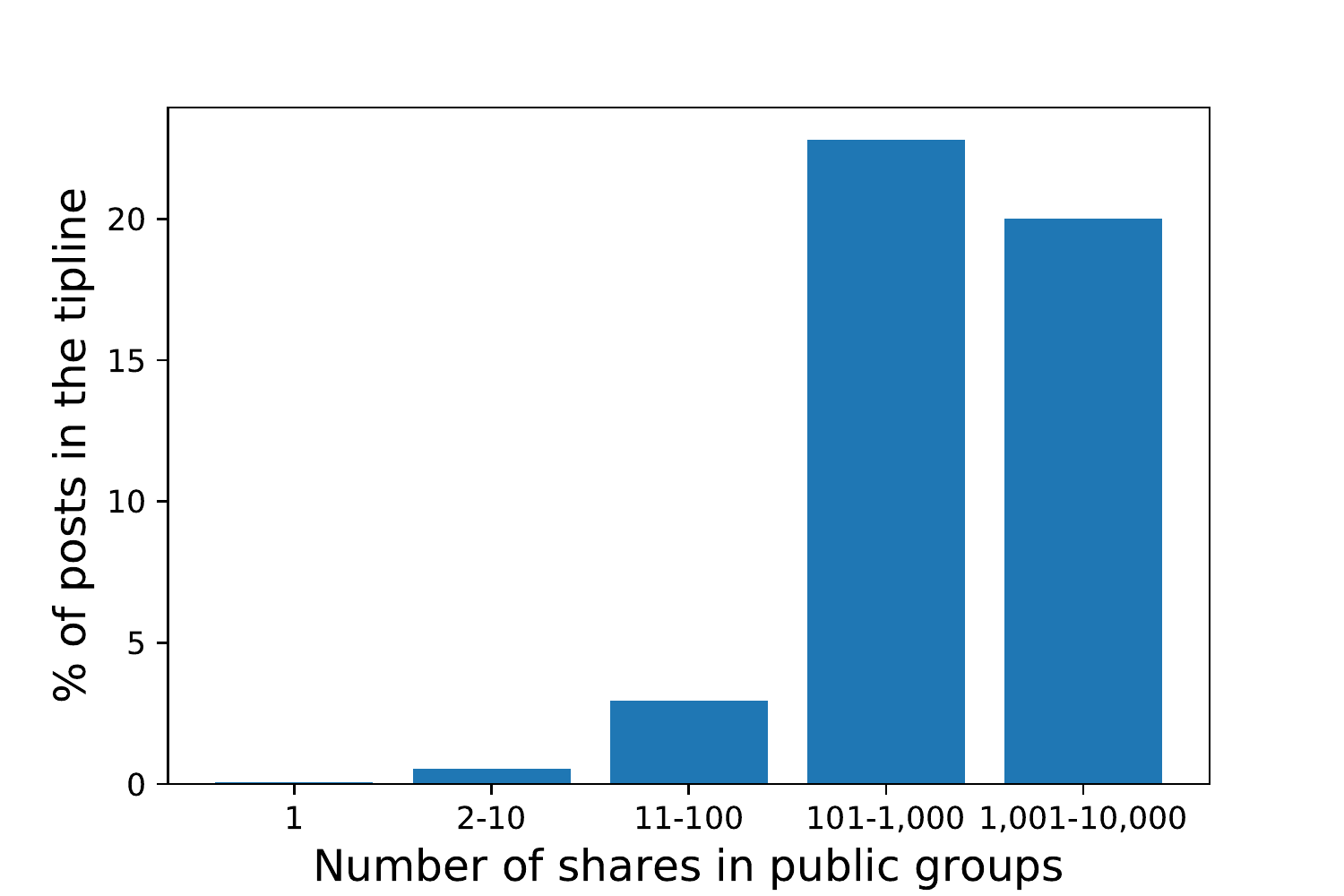}
    \caption{Coverage of text message: The x-axis shows the number of shares on the public groups and y-axis shows the percentage of text message with x shares that match with a text message submitted to the tipline. Text messages that are highly shared on the public groups are much more likely to be also shared to the tipline. Messages in the public groups are first clustered together to determine the number of shares of each message.}
    \label{fig:text-checkpoint-coverage}
\end{figure}

Next, we look at the coverage and compute the number of shares of a piece of content in the public groups or on ShareChat and compute what percentage of the images with different number of shares appear in the tipline dataset.
Figures~\ref{fig:img-checkpoint-coverage} \& \ref{fig:img-checkpoint-sharechat-coverage} show the results. 
For both the public groups and ShareChat, we use logarithmic bucketing of the number of shares of items to estimate message popularity. 
The results show tiplines have good coverage of popular content: 67\% of the images shared 100 or more times in public groups were also submitted to the tipline. We repeat the analysis with text messages and find that 23\% of text messages shared more than 100  times in the public groups were also submitted to the tipline (Figure~\ref{fig:text-checkpoint-coverage}).
To put matters into perspective, we conduct a similar experiment matching all the fact-checked text claims and their corresponding social media posts from the same time period against WhatsApp public groups messages. Only 10\% (12/119) of textual content from viral clusters in public groups (shared more than 100 times) matched with at least one text (claim or fact-checked tweet) from Indian fact-checks during this period.

Exact copies of about 10\% of popular URLs (that is, URLs shared over 1,000 times) on public groups were also submitted to the tipline. Because of shortened URLs, content take-downs, and the 2-year time difference between data collection and analysis, grouping URLs was very challenging. We therefore limit further analysis of URLs for this research question.

We also find text messages and many images submitted to the tipline do not appear in the public groups suggesting tiplines also capture content being distributed in WhatsApp in smaller-group or person-to-person settings.
Out of the 23K unique clusters of images submitted to the tipline, only 5,811 clusters (25\%) had at least one match with an image from the public groups.

Next, we check which \textit{text} messages from the clusters with claims match messages found in the public group data. We find that 93\% of the 257 relevant clusters match at least one message in the WhatsApp public groups dataset. Far from being a skewed result where only few large clusters match, we find a large number of messages across clusters of all sizes match at least one public group message. 
The per-cluster average percent of messages matching to WhatsApp public group data is 91\%. This suggests that if we had included clusters with less than five unique messages, we could have seen additional matches. We did not include these as we only wanted to include messages we knew had fact-checkable claims (and we only annotated clusters with at least five unique messages), but additional annotation could likely yield more relevant messages and matches.

In addition, 7\% of the clusters of fact-check worthy claims from the tipline did not match any public group messages. 
This implies that collecting messages from public group and using tiplines can be complimentary even though neither is a full sample of what is circulating on WhatsApp. 


Finally, we measure the potential impact tiplines could have to prevent the spread of misinformation. 
For this, we looked at items which were shared on both the tipline and in the public groups. We identified the timestamp when an item was first shared on the tipline and counted the number of shares of the item before and after this timestamp on the public groups. The intuition here is that if an item was shared on the tipline, it is in the pipeline to be fact-checked.
The shares after they were shared on the tipline (and hence fact-checked) could have been prevented if we had a mechanism to flag and stop the spread of the images once they were fact-checked.
We find that 38.9\% of the image shares and 32\% of the text message shares in public groups were after the items were submitted to the tipline. 

\subsection{Research Question \#2}
To investigate our second research question, we take an in-depth look into images, text messages, and links sent to the tipline and present examples of the most popular submissions.

\subsubsection{Images}
\begin{figure}[ht]
\centering
\begin{minipage}{.32\linewidth}
\centering
\subfloat[]{\label{a}\includegraphics[width=0.95\textwidth, height=0.95\textwidth]{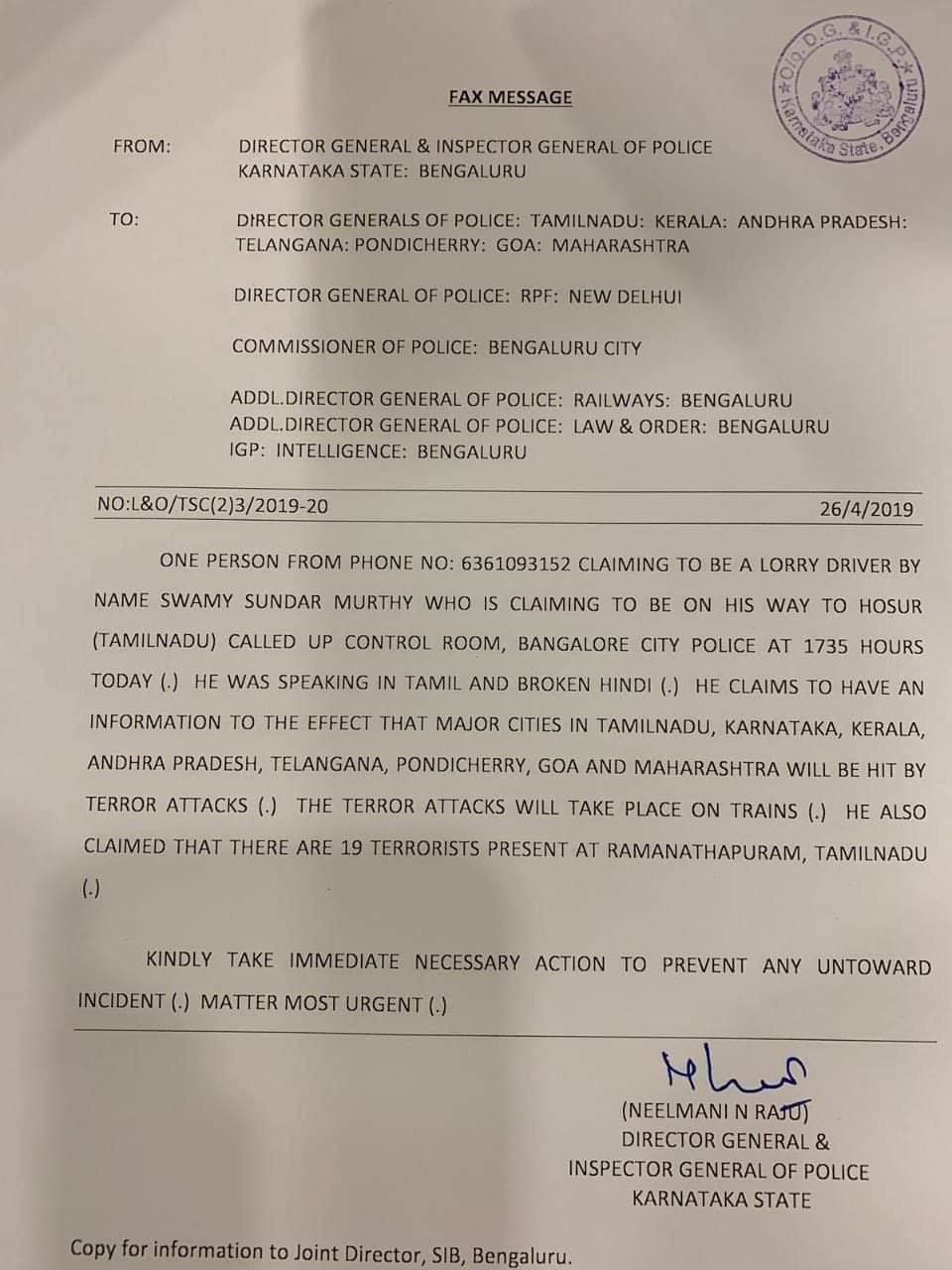}}
\end{minipage}%
\begin{minipage}{.32\linewidth}
\centering
\subfloat[]{\label{b}\includegraphics[width=0.95\textwidth, height=0.95\textwidth]{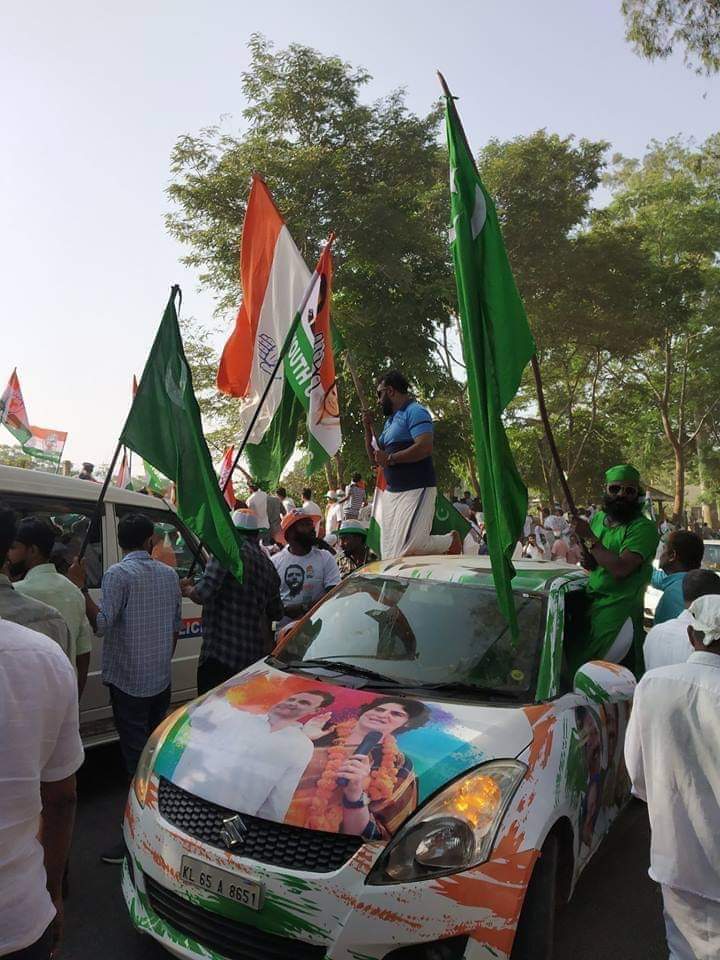}}
\end{minipage}%
\begin{minipage}{.32\linewidth}
\centering
\subfloat[]{\label{c}\includegraphics[width=0.95\textwidth, height=0.95\textwidth]{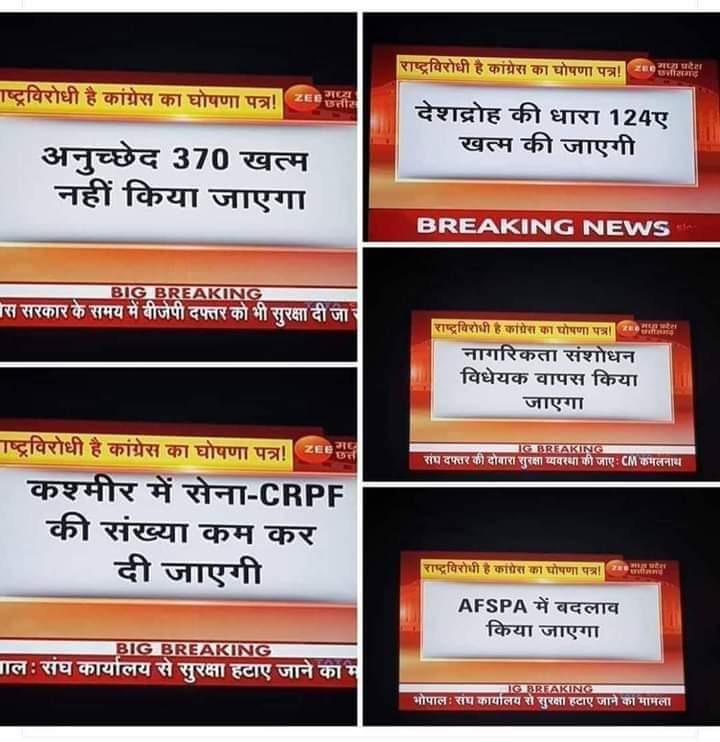}}
\end{minipage}%

\par\medskip
\caption{Most shared images on the tipline. }
\label{fig:checkpoint_popular_images}
\end{figure}

The tipline received 34,655 unique images, which clustered into 23,597 groups. Figure~\ref{fig:checkpoint_popular_images} shows the three most submitted images to the tipline. Each of these three images was submitted by at least 60 unique users. 
All three of these images were fact-checked and found to be false.
Figure~\ref{fig:checkpoint_popular_images}a shows a `leaked' government circular alleging a terrorist plot during the elections. This was just an old circular taken out of context.
Figure~\ref{fig:checkpoint_popular_images}b falsely alleges that Pakistani flags were raised during a political rally, and,
Figure~\ref{fig:checkpoint_popular_images}c shows doctored screenshots of a TV news program.

\begin{figure}[tb]
    \centering
    \includegraphics[width=\columnwidth]{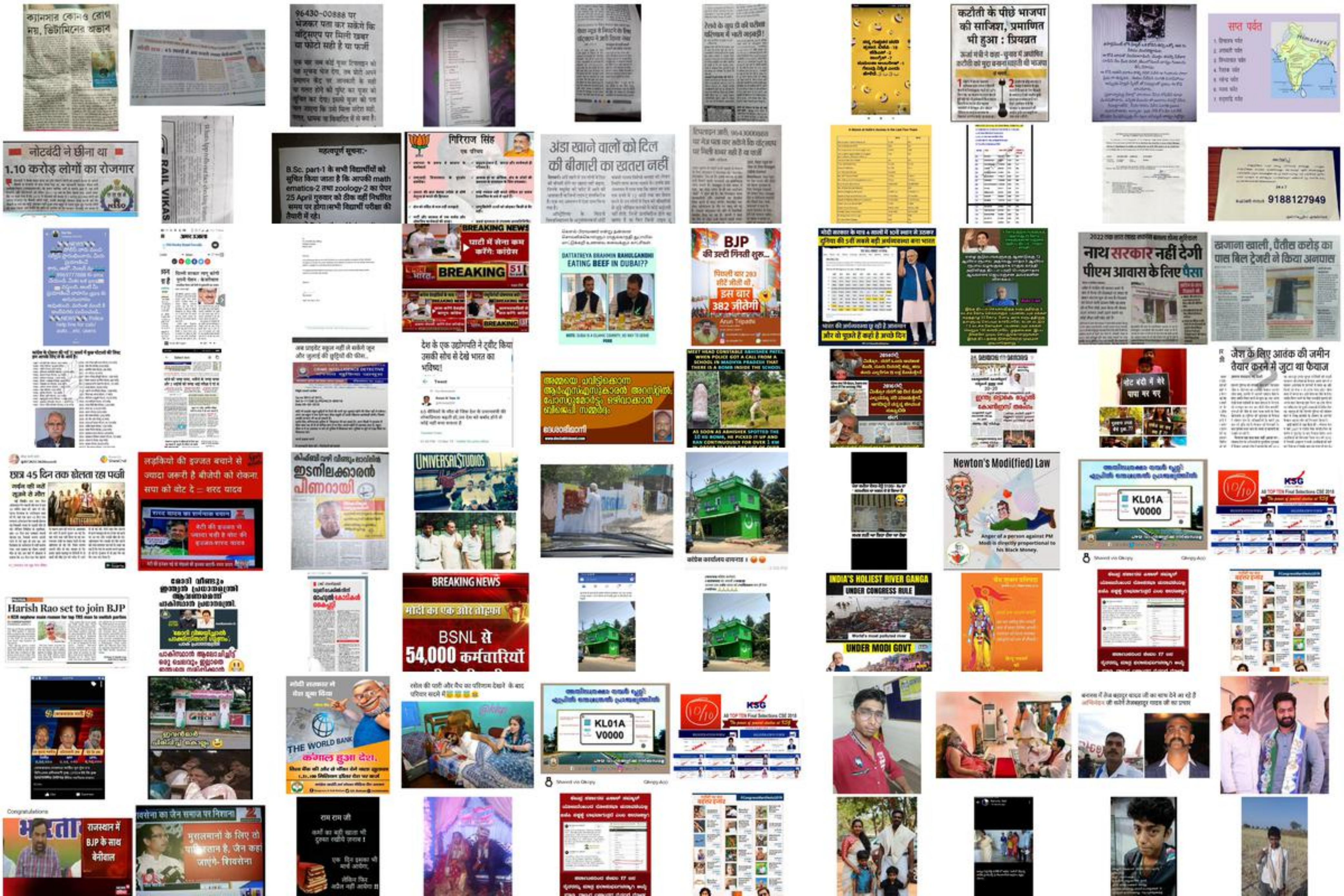}
    \caption{A visual summary of the images submitted to the tipline. The mosaic is a collection of 20 clusters obtained from the 34K images submitted to the tipline. Each cluster is represented as 2x2 grid of images randomly sampled from the cluster.} 
    \label{fig:mosaic}
\vspace{-\baselineskip}
\end{figure}

We construct a visual summary of the 34K images by obtaining a 1,000-dimensional embedding for each image using a pretrained ResNeXt model~\cite{xie2017aggregated}.
Next, we cluster these embeddings using a $k$-means clustering algorithm, and chose $k$=20 using the elbow method.\footnote{\url{https://en.wikipedia.org/wiki/Elbow_method_(clustering)}}
For each cluster, we pick 4 randomly sampled images and create a mosaic of the 20 clusters, shown in Figure~\ref{fig:mosaic}.
The mosaic shows various categories of images sent to the tipline at a high level. 
As we move from the top left to the bottom right, we can see a lot of images on the top left of Figure~\ref{fig:mosaic} containing pictures of newspapers, and in general images with text on it. As we go to the bottom left, we see memes and pictures containing quotes of politicians, and on the bottom right, images of people/politicians.
Pictures of newspapers or images with text on them are the most dominant type, constituting over 40\% of the content, followed by memes which make up roughly 35\% of the content.

\subsubsection{Text messages}

Of the 88K text messages received to the tipline, 37,823 are unique (not exact duplicates). Clustering these messages using the Indian XLM-R model and a threshold of 0.9 results in 20,856 clusters (near duplicates).
There were 559 clusters with five or more unique messages. 
We hired an Indian journalist with experience fact-checking during the Indian 2019 election to annotate each of these clusters for quality of clustering and to identify content that was `fact-check worthy.'~\cite{konstantinovskiy2018towards} The annotation interface presented three examples from each cluster: one with the lowest average distance to all other messages in the cluster, one with the highest average distance to all other messages in the cluster, and one message chosen randomly. We find 257 clusters (out of the 559, 46\%) comprising 2,536 unique messages are claims that could be fact-checked. Overall, 173 clusters (1,945 unique messages, 7,131 total messages) are related to the election, and 84 clusters (591 unique messages, 2,473 total messages) are claims unrelated to the election.

The clusters are generally all high-quality: in 98\% of the clusters all three messages made the same claim. In 2\% of the clusters (11 clusters, 159 unique messages) the three items annotated should not have been clustered. 

There were also 231 clusters that do not have fact-check worthy claims. These were usually advertising/spam (114 clusters, 1,245  unique messages) or messages specific to the tipline (177 clusters, 2,957 unique messages. The tipline-specific messages include messages following up on a previous request, asking for more information about the tipline, and requests for reports in additional languages.

We take the 257 clusters that were annotated as containing claims and find that 203 contain messages in only one language (usually Hindi) while the other clusters contain between two and six languages. Languages were detected via CLD3, and were selected when a known language was detected and that detection was reported as reliable by CLD3.\footnote{\url{https://github.com/google/cld3}}

Within the clusters with election-related claims, the largest cluster is misinformation advising voters to ask for a ``challenge vote'' or ``tender vote'' if they find they are either not on the voter list or have been marked as already voting.\footnote{\url{https://archive.is/BWsqR}} There were 213 unique messages totaling 2,121 submissions to the tipline with this claim across five languages.

Other prominent themes within the election-related clusters include messages attacking BJP leader Narendra Modi, pro-BJP messages, and messages criticizing Congress Party leader Rahul Gandhi.

The largest cluster with a non-election claim is misinformation about the tick marks on WhatsApp claiming that three blue tickmarks indicate the government has observed the message.\footnote{\url{https://archive.is/WfeRe}} 
There were two clusters with different variants of this claim totaling 78 unique messages and 1,000 submissions across Malayalam and English.
Of the 2,536 messages in the clusters containing claims, Hindi (47\%), English (35\%), Malayalam (6\%) are the most common languages. Marathi, Telugu, and Tamil each account for roughly 2\% of the messages.
This likely reflects both the socio-linguistic characteristics of India as well as the fact that the tipline was most heavily advertised in Hindi and English.

In total, there are 9,604 submissions to the tipline for these 2,536 unique messages (that is, 7,068 submissions were exact duplicates).
On average, it took an average of 5 hours (std 1.4, min 0, median 6, max 6) for half of the total number of submissions in each of the clusters with claims to arrive to the tipline. 90\% of the submissions in each cluster arrived within an average of 128 hours (std 17, min 24, median 130, max 153). This suggests slightly slower dynamics than have been seen with the signing of petitions \cite{margetts2015political} and the sharing of news stories on non-encrypted social media \cite{bright2016social}.

\subsubsection{URLs}
Another common content type in WhatsApp groups and tiplines are URLs.
The tipline received 28,370 URLs (12,674 unique URLs), which contain URLs from 2,781 unique domains. A list of most frequent domains is presented in table \ref{tab:domains}. The most prevalent websites submitted to tiplines are social media (\textit{YouTube, Facebook, Twitter and Blogger}), news outlets (IndiaTimes and DailyHunt), and URL shortening services (Bitly and TinyURL). 

\begin{table}[]
\centering
\caption{Top 10 domains most shared with the WhatsApp tipline around the Indian general election period.}
\label{tab:domains}
\begin{tabular}{lc}
\toprule
\textbf{Domain} & \textbf{Total URLs} \\
\midrule
YouTube & 2,350 \\
Blogger & 2,107 \\
Bitly & 1,636 \\
Google & 1,471 \\
Facebook & 1,192 \\
RechargeLoot & 724 \\
IndiaTimes & 587 \\
DailyHunt & 574 \\
Twitter & 515 \\
TinyURL & 465 \\
\bottomrule
\end{tabular}
\end{table}

\section{Methods \& Data}
\subsection{Methods}
\subsubsection{Image Similarity.}
To identify similar images we use Facebook's PDQ hashing and Hamming distance. PDQ is a perceptual hashing algorithm that produces 64-bit binary hash for any image. Small changes to images result in only small changes to the hashes and thus allow visually similar images to be grouped. This allows, for instance, the same image saved in different file formats to be identified. For this paper, images with a Hamming distance of less than 31 are considered to be similar. The same threshold was used previously by~\citet{reis2020ondevice}.
Similar images were clustered together using the DBSCAN~\cite{ester1996density} algorithm.

\subsubsection{Text Similarity.}
To identify similar textual items, we use a multilingual sentence embedding model trained for English, Hindi, Bengali, Marathi, Malayalam, Tamil, and Telugu \cite{kazemi2021claim}. 
\citet{kazemi2021claim} evaluated this model for claim matching using similar data and found applying a cosine similarity threshold of 0.9 to pairs of messages resulted in the best performance, with an overall F1 score of 0.73. The model performs better on English and Hindi (which make 82\% of the data), with an average F1 score of 0.85.
Throughout this paper we use cosine similarity and a threshold of 0.9 for matching text items.

\subsubsection{Text Clustering.}
To compare within and between tiplines, we cluster text items using online, single-link hierarchical clustering. Each new message arriving to the tipline is compared to all previous messages, and the best match found. If this match is above the similarity threshold, then we add the new message to the same cluster as the existing message. We apply the same process to the public group messages.
To enable quick retrieval, we construct a FAISS~\cite{douze2017faiss} index using our Indian XLM-R embeddings of all the public group messages. We then query this index for each tipline message and record all matches with a cosine similarity score of at least 0.9. We remove any duplicate matches (i.e., cases where two tipline messages matched the same public group message) before analyzing the matches.

\subsection{Data}

We used a wide range of data sources in this work, from the WhatsApp tipline data, to social media data from WhatsApp public groups and ShareChat, and published fact-checks. All the data used pertained to the four month period between March 1, 2019, and June 30, 2019. This period includes the 2019 Indian general election, which took place over a period of 6 weeks in April and May 2019.

\subsubsection{Tiplines.}
In 2019, PROTO led the Checkpoint project using Meedan's open-source software to operate a WhatsApp tipline. PROTO advertised their WhatsApp number asking users to forward any potentially misleading content related to the election. They advised that they would be able to check and reply to some of the content that they received. Over the course of four months, 157,995 messages were received. Of these, 82,676 were unique and consisted of 37,823 text messages, 10,198 links, and 34,655 images.

We obtained a list of links, text messages, and images along with the timestamps of when they were submitted to the tipline. We have no information about the submitting users beyond anonymous ids.

\subsubsection{WhatsApp Public Groups.}
There are currently over 400 million active WhatsApp users in India. With the availability of cheap internet data and smartphones with WhatsApp pre-installed, the app has become ubiquitous. Aside from messaging friends and family, Indians use WhatsApp to participate in political discourse \cite{farooq2017politics}. Political parties have taken this opportunity to create thousands of public groups to promote their political agenda. These groups have been shown to be quite prevalent, with over 1 in 6 Indian WhatsApp users being a part of one such group~\cite{lokniti2018}. 

In addition to the image and text items submitted to tiplines, we have data from large ``public'' WhatsApp groups collected by \citet{garimella2020images} during the same time period as the tipline run by PROTO. 
The dataset was collected by monitoring over 5,000 public WhatsApp groups discussing politics in India. For more information on the dataset, please refer to \citet{garimella2020images}.

\subsubsection{ShareChat.}
ShareChat is an Indian social network which is used by over 100 million users.\footnote{\url{https://sharechat.com/}} It has features similar to Instagram and is primarily multimedia focused~\cite{agarwal2020characterising}. 
Unlike WhatsApp, ShareChat provides global popularity metrics including likes and share count, which allow us to construct a proxy for the popularity of the content on social media.
ShareChat curates popular hashtags based on topics such as politics, entertainment, sports, etc. During the three months of data collection, every day, we obtained the popular hashtags related to politics and obtained all the posts containing those hashtags.
This provides a complete sample of all images related to politics which were posted on ShareChat during the data collection period (March 1 to June 30 2019).

\subsubsection{Fact-checks.}
We also collect fact-checks and social media data from the time period in English and Hindi. We crawled popular fact-checking websites in India and obtained articles and any tweets linked within the articles following the approach of \citet{shahi2020amused} and \citet{shahi2021exploratory}. Overall, we find 18,174 fact-check articles in 49 languages from 136 fact-checkers from all over the globe. To get the fact-checked article concerning the Indian General election, we filtered the data based on the fact check being in an Indian language or the fact-checking domain being based in India.

In total, we obtained 3,224, and 2,220 fact-checked articles in English and Hindi respectively. 
The articles fact-check content from various social media platforms, including Twitter. Whenever available, we also obtained the links to the original tweets that were fact-checked and downloaded these. We obtained 811 tweets in total, 653 (182 unique) in English and 158 (63 unique) in Hindi.

A summary of all the data collected is shown in Table~\ref{tab:data}.

\begin{table}[]
\centering
\small
\caption{Datasets used in this work. The values shown in parentheses indicate the number of unique messages/images. We only collected image data from ShareChat.}
\label{tab:data}
\begin{tabular}{llll}
\toprule
\multicolumn{1}{l}{\textbf{Datasets}} & \multicolumn{1}{l}{\textbf{\begin{tabular}[c]{@{}l@{}}\#Text messages\\ (unique)\end{tabular}}} & \multicolumn{1}{l}{\textbf{\begin{tabular}[c]{@{}l@{}}\#Images\\ (unique)\end{tabular}}} \\
\midrule
Public groups & 668,829 (445,767) & 1.3M (977K)\\
ShareChat & - & 1.2M (401K)\\
Checkpoint & 88,662 (37,823) & 48,978 (34,655) \\ 
Fact-check articles & 5,444 (5,444) & - \\
Fact-check tweets & 811 (245) & - \\
\bottomrule
\end{tabular}
\end{table}

\section{Acknowledgments}
We are grateful to the Meedan team, Pop-Up Newsroom, PROTO, and Prof.\ Rada Mihalcea for valuable feedback and data access.

\section{Funding}
This work was funded by the Omidyar Network with additional support from Sida, the Robert Wood Johnson Foundation, and the Volkswagen Foundation. 
Kiran Garimella is supported by the Michael Hammer postdoctoral fellowship at MIT.

\section{Competing interests}
Meedan is a technology non-profit that develops open-source software that many fact-checking organizations use to operate misinformation tiplines on WhatsApp and other platforms. The research team at Meedan operates independently: the research questions, approach, and methods used in this article were decided by the authors alone. 

\section{Ethics}
Throughout this research, we were extremely concerned about the privacy of tipline users and the ethical concerns that come with large-scale studies. All WhatsApp messages in our datasets were anonymized, and personal information (e.g., phone numbers) was removed. Our experiments were done at the macro level, and we followed strict data access, storage, and auditing procedures to ensure accountability.



\bibliographystyle{apacite}
\bibliography{refs}
\end{document}